\documentclass{iaus}
\usepackage{graphicx}

\title[Model of torsional oscillations] 
{A theoretical model of torsional oscillations from a flux transport dynamo model}

\author[Chatterjee, Chakraborty \& Choudhuri]   
{Piyali Chatterjee$^1$,
Sagar Chakraborty$^2$
 \and Arnab Rai Choudhuri$^3$}

\affiliation{$^1$NORDITA, AlbaNova University Center, Roslagstullsbacken 23,  
\\ SE 10691 Stockholm, Sweden \\ email: {\tt piyalic@nordita.org} \\[\affilskip]
$^2$NBIA, Niels Bohr Institute, Blegdamsvej 17, 
\\ 2100 Copenhagen, Denmark \\email: {\tt sagar@nbi.dk} \\[\affilskip]
$^3$Department of Physics, Indian Institute of Science, \\
Bangalore -- 560012, India \\email: {\tt arnab@physics.iisc.ernet.in}}

\pubyear{2010}
\volume{}  
\jname{Model of torsional oscillations}
\editors{Debi Prasad Choudhary, Klaus G. Strassmeier eds.}

\def\pa{\partial}
\def\vp{v_{\phi}}

\begin{document}

\maketitle

\begin{abstract}

Assuming that the torsional oscillation is driven by the Lorentz
force of the magnetic field associated with the sunspot cycle, we
use a flux transport dynamo to model it and explain its initiation
at a high latitude before the beginning of the sunspot cycle.
\end{abstract}

\firstsection 
\section{Introduction}

The small periodic variation in the Sun's rotation with the
sunspot cycle, first discovered on the solar surface by Howard
\& LaBonte (1980), is called torsional oscillations.
Helioseismology has now established its existence throughout
the convection zone (see Howe et al.\ 2005 and references
therein). Its amplitude
near the surface is of order 5 m s$^{-1}$ or about 1\% of
the angular velocity. Apart from
the equatorward-propagating branch which moves with the sunspot
belt, there is also a 
poleward-propagating branch at high latitudes. One intriguing
aspect of the equatorward-propagating branch is that it
begins a couple of years before the sunspots of a particular
cycle appear and at a latitude higher than where
the first sunspots are seen.  The top panel of Fig.~1 shows
the torsional oscillations at the solar
surface with the butterfly
diagram of sunspots. If the torsional oscillation
is caused by the Lorentz force of the dynamo-generated
magnetic field as generally believed, 
then the early initiation of this oscillation
at a higher latitude does 
look like a violation of causality! Our main aim is to explain
this which could not be explained by the earlier theoretical
models (Durney 1980; Covas et al.\ 2000; Bushby 2006; Rempel
2006).  The details of our work can be found in a recent paper
(Chakraborty, Choudhuri \& Chatterjee 2009a, hereafter CCC).  
Please note that this paper has an 
erratum (Chakraborty, Choudhuri \& Chatterjee 2009b).

\section{Theoretical model}

The flux transport dynamo model first developed by Choudhuri,
Sch\"ussler \& Dikapti (1995) appears to be the most promising
model for explaining the sunspot cycle.  We use the model presented
by Chatterjee, Nandy \& Choudhuri (2004).  Some details of the
model with the basic equations can be found in Choudhuri (2010).
In order to model torsional oscillations, in addition to the
basic equations of the dynamo, we simultaneously have to solve
the Navier--Stokes equation in the form
$$\rho \left\{ \frac{\pa \vp}{\pa t} + D_v [\vp] \right\} = 
D_{\nu} [\vp] + ({\bf F}_L)_{\phi},\eqno(1)$$ 
where 
$D_v [\vp]$ is the term corresponding to advection by the meridional circulation,
$D_{\nu} [\vp]$ is the diffusion term, and 
$({\bf F}_L)_{\phi}$ is the $\phi$ component of the Lorentz force.
\def\Bp{B_{\phi}}
If the magnetic field is assumed
to have the standard form
$${\bf B} = B (r, \theta, t){\bf e}_{\phi} + \nabla \times 
[A(r, \theta, t){\bf e}_{\phi}], \eqno(2) $$
then the Lorentz force is given by the Jacobian
$$4 \pi ({\bf F}_L)_{\phi} = \frac{1}{s^3} J \left( \frac{s B_{\phi}, 
s A }{r, \theta} \right), \eqno(3)$$
where $s= r \sin \theta$. On the basis of flux tube simulations
suggesting that the magnetic field in the tachocline should be
of order $10^5$ G (Choudhuri \& Gilman 1987; Choudhuri 1989;
D'Silva \& Choudhuri 1993), it is argued by Choudhuri (2003)
that the magnetic field has to be
intermittent in the tachocline. Hence the full expression
of Lorentz involves a filling factor as explained
by CCC.

\begin{figure}
\center
\includegraphics[width=10cm]{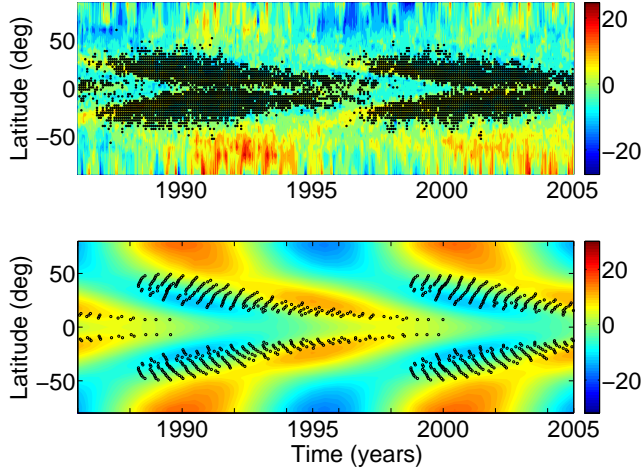}
\caption{The time-latitude plot of torsional oscillation on the
solar surface with the butterfly diagram of sunspots superposed
on it. The upper panel is based on observational data of surface
velocity $\vp$ measured at Mount Wilson Observatory (courtesy:
Roger Ulrich).  The bottom panel is from our theoretical simulation.}
\end{figure}

Our theoretical model incorporates a hypothesis proposed by
Nandy \& Choudhuri (2002), which is essential for explaining the
early initiation of the torsional oscillation at high latitudes.
According to this Nandy--Choudhuri (NC) hypothesis, the
meridional flow penetrating in stable layers below convection 
zone causes formation of toroidal field in high latitude tachocline. 
Sunspots form a few years later when this field is advected to 
lower latitudes and brought inside convection zone. We also
assume that the stress of the magnetic field formed in the tachocline
is carried upward by Alfven waves propagating along vertical flux
concentrations conjectured by Choudhuri (2003).

\begin{figure}
\begin{minipage}[b]{0.5\textwidth}
\includegraphics[height=4cm,width=6cm]{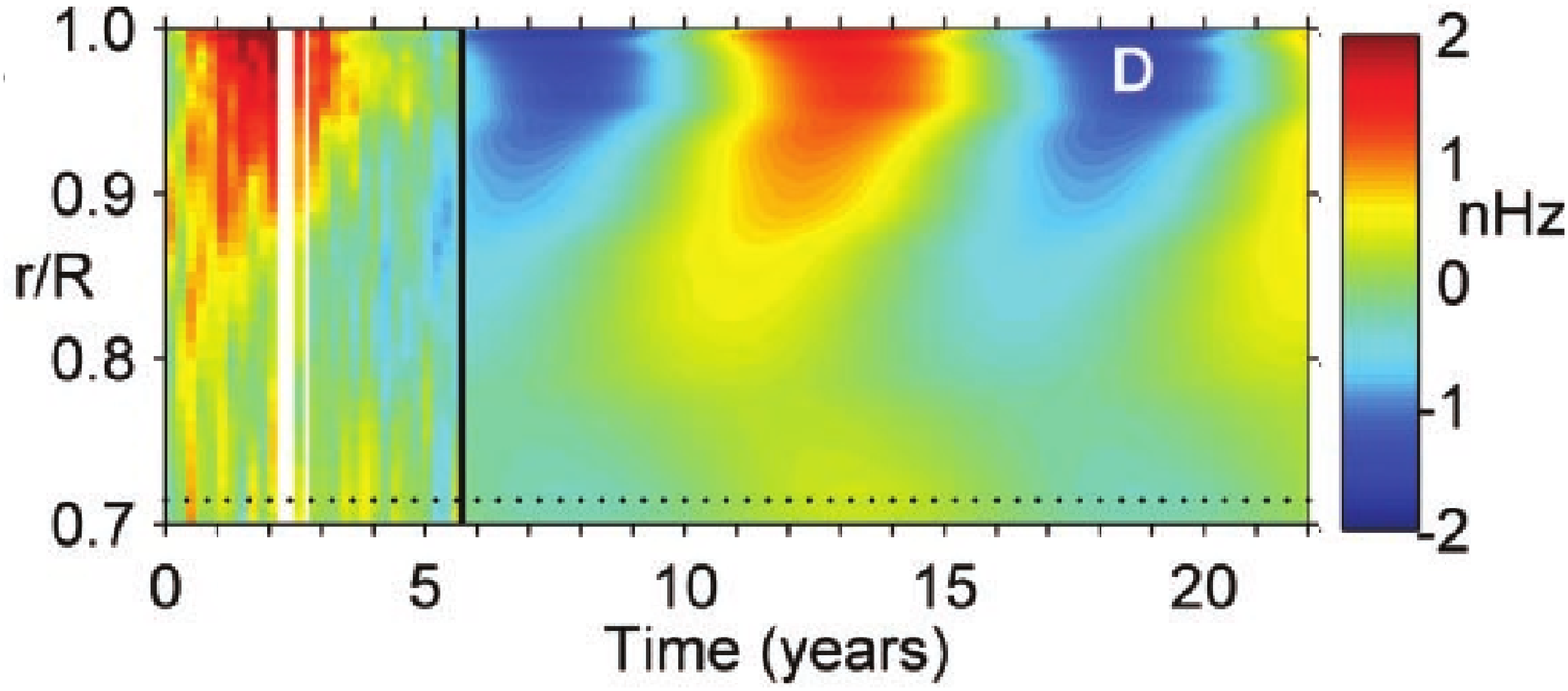}
\end{minipage}
\begin{minipage}[b]{0.5\textwidth}
\includegraphics[height=4cm,width=6cm]{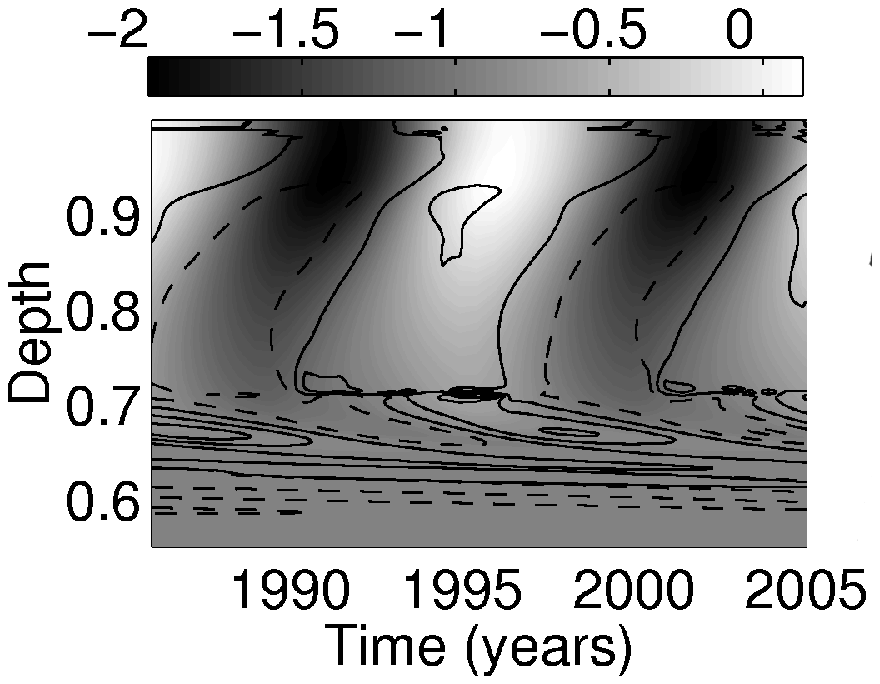}
\end{minipage}
\caption{The depth-time plot of torsional oscillations
at latitude $20^{\circ}$.
The left panel from Vorontsov et al. (2002)  
is based on SOHO observations,
whereas the right panel from CCC is based
on our theoretical simulation.
The solid and dashed lines in the right panel indicate the Lorentz force
(positive and negative values respectively).}
\end{figure}

\section{Results of simulation}

The incorporation of the NC hypothesis in our theoretical model
causes magnetic stresses to build up at higher latitudes before 
sunspots of the cycle appear, leading to the early 
initiation of torsional oscillations.
The bottom panel of Fig.~1 shows theoretical results of torsional
oscillations at the surface with the theoretical butterfly diagram.
This bottom panel can be compared with the observational upper
panel in Fig.~1. Our theoretical model also gives a satisfactory
account of the evolution of torsional oscillations within the convection
zone.  The depth-time plot of torsional oscillations at a certain
latitude given in Fig.~3 of CCC compares favourably with the 
observational plot given in Fig.~3(D) of Vorontsov et al.\ (2002). 
This is reproduced in Fig.~2 for completeness.

\end{document}